\documentclass{llncs}
\usepackage[usenames]{color}
\usepackage{epsf}
\usepackage{subfigure,floatflt,wrapfig}
\usepackage{graphicx}
\graphicspath{
{figures/}
}

\begin{document}
\title{Modular RADAR: An Immune System Inspired Search and Response Strategy for Distributed Systems}
\author{Soumya Banerjee\inst{1} \and Melanie Moses\inst{1}}
\institute{Department of Computer Science, University of New Mexico, USA \email{\{soumya,melaniem\}@cs.unm.edu}}

\maketitle

\begin{abstract}
The Natural Immune System (NIS) is a distributed system that solves challenging search and response problems while operating under constraints imposed by physical space and resource availability. Remarkably, NIS search and response times do not scale appreciably with the physical size of the animal in which its search is conducted. Many distributed systems are engineered to solve analogous problems, and the NIS demonstrates how such engineered systems can achieve desirable scalability. We hypothesize that the architecture of the NIS, composed of a hierarchical decentralized detection network of lymph nodes (LN) facilitates efficient search and response. A sub-modular architecture in which LN numbers and size both scale with organism size is shown to efficiently balance tradeoffs between local antigen detection and global antibody production, leading to nearly \textit{scale-invariant detection and response}. We characterize the tradeoffs as balancing local and global communication and show that similar tradeoffs exist in distributed systems like LN inspired artificial immune system (AIS) applications and peer-to-peer (P2P) systems. Taking inspiration from the architecture of the NIS, we propose a modular RADAR (Robust Adaptive Decentralized search with Automated Response) strategy for distributed systems. We demonstrate how two existing distributed systems (a LN inspired multi-robot control application and a P2P system) can be improved by a modular RADAR strategy. Such a sub-modular architecture is shown to balance the tradeoffs between local communication (within artificial LNs and P2P clusters) and global communication (between artificial LNs and P2P clusters), leading to efficient search and response.
\end{abstract}

\section{Introduction}

Distributed systems are becoming increasingly important, for example in environmental monitoring, disaster relief, military operations, multi-robot control, wireless sensor networks and peer-to-peer systems \cite{1}. Such systems typically are distributed in physical space with limited resources like power and bandwidth, and performance scalability is desirable. The natural immune system (NIS) demonstrates that nearly \textit{scale invariant} performance is possible, even under constraints imposed by physical space and resource availability \cite{2,3}. 

The NIS searches trillions of host cells to find small amounts of spatially localized antigen. The search for antigen and response (neutralizing the pathogen) is nearly invariant with respect to the size of the host \cite{2,3}. Search in the NIS occurs in a hierarchical modular network of lymph nodes (LN) distributed throughout the body. LNs are the primary locations in which populations of adaptive and innate immune cells interact to recognize and respond to pathogens. We hypothesize that this NIS architecture promotes modular RADAR (Robust Adaptive Decentralized search with Automated Response). 

Previous work \cite{2,3} shows how the architecture of the lymphatic network enables the NIS to detect antigen and respond by producing antibodies in time that is nearly invariant with animal size. Our analysis of data on experimental infection by West Nile Virus (WNV) of a range of animals (from 30g sparrows to 300 kg horses) indicates that the NIS neutralizes WNV in approximately 3 days regardless of body size \cite{2,3}. This is surprising in the light of two facts: First, the NIS of a horse has to search for a ÒneedleÓ (rare antigen) in a much larger ÒhaystackÓ (body) than the NIS of a sparrow. Second, experimental data indicates that a critical concentration of WNV-specific antibody needs to be present in the bloodstream to neutralize WNV. Since the volume of blood in an organism is proportional to body mass \cite{4}, a horse will have to produce 10000 times more antibody, in the same time, as a sparrow.   

Hence the search and response problem is more challenging in larger organisms. We show that a semi-modular NIS architecture is capable of nearly \textit{scale-invariant detection and response}. We then use this as inspiration for search and response strategies in other distributed systems. We show how the modular RADAR architecture improves search and response in a LN inspired multi-robot control application and a peer-to-peer content search system. 

The remainder of the paper is organized as follows: we review relevant features of the NIS, define attributes of the modular RADAR process, explain our experimental data source, and then explain how a sub-modular architecture makes scale-invariant search and response possible. We then demonstrate how modular RADAR architectures improve search in two distributed systems, and end with concluding remarks. 

\section{A Review of the Relevant Immunology}

LNs reduce the physical space in which lymphocytes search for the particular pathogens that they can recognize and neutralize, and hence they vastly speed up the distributed search and response to novel pathogens. Without concentrating the search into a small volume of tissue, the distributed search for novel pathogens by rare pathogen
specific lymphocytes would be prohibitively slow.

We hypothesize that evolutionary pressures have shaped the number and size of LN to minimize the time to unite rare antigen-specific lymphocytes (B-cells and T-cells) with pathogens. LNs provide a small volume of tissue in which lymphocytes, antigen-presenting cells and antigen encounter each other. The area of tissue that drains into a LN is called its draining region (DR). Dendritic cells (DCs) sample the tissue in DRs for pathogens, and upon encountering them, migrate to the nearest LN to present antigen to T helper cells. Upon recognizing cognate antigen on DCs, T-cells proliferate and build up a clonal population in a process called clonal expansion. While proliferating, T helper cells also migrate to the LN B-cell area to activate B-cells. B-cells are activated by antigen-bearing DC and T helper cells. After activation, B-cells undergo clonal expansion and differentiate into antibody-secreting plasma cells.

Cognate T-cells and B-cells specific to a particular pathogen are very rare ($1$ in $10^{6}$ cells will bind to a particular antigen \cite{5}), thus the draining LN also recruits B-cells from other LNs. The rate at which B-cells flow into a particular LN is determined by the cross sectional area of the High Endothelial Venule (HEV) which is proportional to LN volume \cite{5}. Rapid detection requires that pathogens are quickly picked up by DC and carried to the LN for presentation to T- and B-cells. Smaller DR reduce the time for DC to migrate to the LN. Rapid response (in this case we consider antibody production as the response), requires that sufficient numbers of B-cells are activated, ultimately to produce sufficient antibody to neutralize the pathogen. Since both the number of B-cells resident in a LN and the rate at which B-cells are recruited to a LN is proportional to LN volume, larger LN reduce the time to activate a B-cell response.

\section{Modular RADAR}

LNs form a hierarchical distributed detection network that is modular (by dividing the search space into semi-independent modules or LNs), efficient (since search occurs in the small physical space of a LN as opposed to throughout the body) and parallel (since there are many such small searches being conducted in parallel). The size of a module (LN) is set by tradeoffs between local and global communication. We mathematically determine the extent of modularity required to conduct efficient and nearly scale-invariant search and response. We call this a modular RADAR (Robust Adaptive Decentralized search with Automated Response) strategy. Search and response in the NIS is characterized by the following attributes:

\begin{enumerate}
\item Robust: There is redundancy and diversity among components, but the exact role of a component is flexible. Large numbers of individuals respond probabilistically to information signals. While sometimes this can lead to responses that grow out of control (i.e. cytokine storms in response to influenza infection \cite{6}), generally the \textit{wisdom of the crowd} \cite{7} is effective even when individuals make mistakes. 
\item Adaptive: Adaptation occurs by individuals or populations of individuals changing in response to environmental signals. When a B-cell binds to a pathogen and is activated by other immune system cells, that B-cell produces a large and variable population of daughter B-cells. Those that bind to the pathogen most effectively reproduce faster, so the population of cells gets better at neutralizing the pathogen.
\item Decentralized search: Control is completely decentralized and communication between NIS cells is aggregated spatially, e.g. LNs concentrate interactions and chemical signalling between NIS cells.
\item Automated Response: The response is as distributed as search. Individuals act (e.g., kill an infected cell or move along a chemical gradient) by integrating local signals from their environment.
\end{enumerate}

Much of the Artificial Immune System (AIS) literature focuses on abstracting algorithms to confer adaptation, learning, memory and robustness from NIS processes (e.g. negative selection, clonal selection algorithms, etc \cite{8}). We do not suggest new algorithms; rather, we propose that the physical division of NIS search into LN and DR  can inspire architectures that other distributed algorithms can exploit to balance local and global communication.

\section{Data Characterizing Immune Response Times}
In previous work \cite{2,3}, we analyzed data from a study of WNV infection. Animals ranging from 0.03 kg sparrows to 300 kg horses, were experimentally infected with WNV and the viral concentration in blood was measured on a daily basis. A critical amount of WNV-specific antibody is required for neutralizing WNV \cite{9}. Previous studies have shown that the time to attain peak viral concentration ($t_{pv}$) coincides with the time taken to secrete a critical amount of WNV-specific antibody \cite{9}. For the experimental infection studies described above, $t_{pv}$ ranged from 2 to 4 days post infection, and was uncorrelated with animal body size ($t_{pv}$ vs. $M$, p-value = 0.35, where $M$ is organism mass) \cite{2,3}. Since $t_{pv}$ is the time taken to detect pathogens and secrete a critical amount of antibody, the absence of a relationship of $t_{pv}$ with mass indicates that the NIS has nearly scale-invariant detection and response. 

\section{Sub-Modular Architecture Balances Tradeoff Between Local and Global Communication}
\subsection{Data on Lymph Nodes}

Published empirical data suggest that the mammalian NIS has a sub-modular architecture. LNs increase in both size and in numbers as animal size increases, e.g. 20g mice have 24 LN averaging 0.004g each, and humans are 3000 times bigger and have 20 times more LN, each 200 times bigger \cite{10,11}. Data from elephants (with LN approaching the size of an entire mouse) and horses (with 8000 LN) also support the hypotheses that LN size and number both increase with body size \cite{11,12}; however data for more species are required in order to determine scaling exponents that quantitatively describe how LN size and number change with body size.

We hypothesize that total LN volume (the number of LN multiplied by average LN volume) is linearly proportional to animal mass because most organ and fluid volumes (i.e heart, blood and liver volume) are linearly proportional to mass \cite{4}. The data are consistent with this hypothesis. This suggests a tradeoff between LN size and number. Since DR size equals animal volume divided by LN number, there is also a tradeoff between DR size and LN size.

\subsection{Mathematical Analysis}

In previous work \cite{2,3} we hypothesized that the NIS is sub-modular because it is selected not just to minimize time to detect pathogens, but also to minimize the time to produce a sufficient concentration of antibody in the blood ($Ab_{crit}$). We hypothesize that the NIS has evolved to minimize two quantities: the time to detect antigen and the time to recruit B-cells from other LNs and produce absolute quantities of antibody ($Ab$), where $Ab$ is proportional to $M$ (organism mass). 

The smallest LN (e.g. in mice or sparrows) contain on the order of a single B-cell that recognizes any particular pathogen \cite{5}. If LNs in all organisms were of a fixed size (and therefore contained a fixed number of B-cells), each LN would serve a fixed-size DR of tissue, and a parallel search for antigen in a fixed space would be repeated more times in a larger organism. This would achieve scale-invariant search. 

However, the NIS of larger organisms has to activate a number of B-cells ($B_{crit}$) $\propto M$, in order to build up the critical density of antibodies in a fixed period of time \cite{9}. Activating $B_{crit}$ (which is proportional to $M$) to fight an infection like WNV that is initially localized in a single DR, requires recruiting on the order of $M$ B-cells from distant LN. We consider this activation of B-cells from remote LN as global communication.  

The number of LNs that a single infected site LN has to communicate with ($N_{comm}$) in order to recruit more B-cells is proportional to the critical number of B-cells required to neutralize the pathogen ($B_{crit}$) divided by the number of antigen-specific B-cells resident in a LN ($Num_{Bcell}$):  $N_{comm} \propto B_{crit}  / Num_{Bcell}$.  Noting that $B_{crit} \propto M$ and $Num_{Bcell} \propto V_{LN}$  (volume of a LN) we have $N_{comm} \propto M / V_{LN}$.   

The rate at which new B-cells from other LN enter into infected LN through the high endothelial vessels ($rate_{comm}$) is proportional to the volume of the LN, $rate_{comm} \propto V_{LN}$ \cite{5}. The time spent in communicating with other LNs and recruiting and activating other B-cells ($t_{comm}$) is then given by $t_{comm} = N_{comm} / rate_{comm}$ and $t_{comm} \propto M / V_{LN}^{2}$.	
   
Hence if LNs in all organisms are of the same size (if $V_{LN} \propto M^{0}$), and the number of LNs scaled linearly with organism size, there are increasing costs to communicating with other LN as the organism gets bigger ($t_{comm} \propto M$); the NIS would conduct efficient search but not efficient antibody production. 

An alternative is to have a fixed number ($N$) of LN whose volumes increase linearly with $M$: $V_{LN} \propto M^{1}$ and $N \propto M^{0}$. This increases the rate of influx of B-cells ($rate_{comm} \propto V_{LN} \propto  M$) and also situates more NIS cells inside the infected site LN. Since all the necessary NIS cells which need to be activated are within the LN, this architecture has no communication cost. However, if the number of LN is invariant, each LN in a larger organism services a larger volume of tissue. Such an architecture would lead to DC migration times that are prohibitively long for large animals ($t^{DC}_{migrate} \propto M^{1/3}$) since antigen-loaded DCs would have to migrate longer distances to reach the draining LN [3].

The architecture that strikes a balance between the two opposing goals of antigen detection (local communication) and antibody production (global communication) is found by minimizing $t_{comm}$ and $t_{migrate}$ which gives $V_{LN} \propto M^{3/7}$ and $N \propto M^{4/7}$ where $V_{LN}$ is the size (volume) of a LN and $N$ is the number of LNs in an organism of mass $M$ \cite{3}. This is a sub-modular architecture in which LN size and numbers both scale sub-linearly with organism size (Fig. 1). 

These equations indicate that time for DC to carry antigen to LN scales as $M^{1/7}$ (since $t_{migrate}$ is proportional to the radius of the DR, whose volume is $M/N$ or $M^{3/7}$). Similarly, the time to recruit cognate B-cells to the infected site is $N$ divided by the rate of B-cell migration into the LN (which is proportional to $V_{LN}$), giving, again, $M^{1/7}$. In previous work \cite{3} we use an agent based model parameterized with empirical values to show that this scaling results in a total time for the entire search and response process that differs by less than 1 day between horses and sparrows (which is in agreement with empirical data).

In summary, due to the requirement of activating increasing numbers of NIS cells for antibody production in larger organisms, there are increasing costs to global communication as organisms grow bigger. The sub-modular architecture balances the opposing goals of detecting antigen using local communication and producing antibody using global communication. Such a modular RADAR search strategy leads to optimal antigen detection and antibody production time, and makes the NIS robust and adaptive. 

\begin{figure}[ht!]
 \includegraphics[width=1\textwidth]{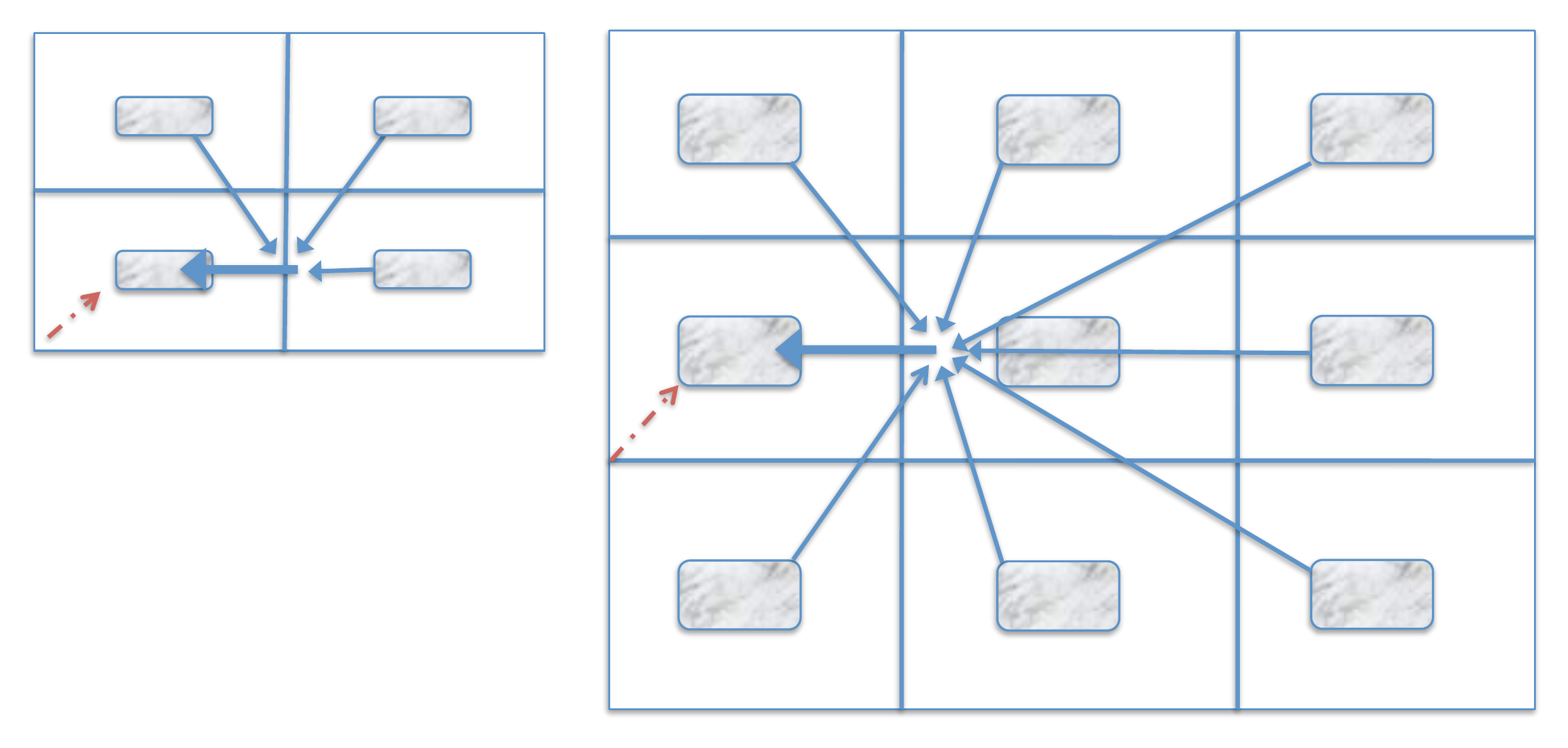}
\caption{A sub-modular detection network of lymph nodes. The shaded regions are LNs and the unshaded regions are the DR. The hypothetical organism on the right is four times as big as the one on the left. The number of LNs and their size both increase with the size of the organism. The local communication (antigen loaded DC migration to LN) is shown by a dotted arrow and the global communication (recruitment of NIS cells from other LNs) is shown by solid arrows. The size of the incoming arrow into a LN represents the size of the HEV (which is proportional to the size of a LN)} 
 \label{fig:Fig1}
\end{figure}

\section{Applications to Artificial Immune Systems and Other Distributed Systems}

Spatially constrained networks are being increasingly used in environmental monitoring, disaster relief and military operations \cite{1}. These networks operate under constraints similar to an NIS and hence their design can be informed by architectural strategies employed by their biological counterpart. We incorporate modular RADAR strategies for rapid search and response to two distributed systems.

\subsection{Application to Mobile Devices}

\subsubsection{Original System}
Our work has implications for the architecture of AIS algorithms for multi-robot control \cite{13,14,15}. The robots and their communication are constrained by physical space, and scaling of performance with system size is an important design criterion. These systems use computer servers (analogous to LNs) to coordinate information exchange between mobile devices (analogous to dendritic cells) in a physical area (analogous to the DR of a LN). When the robots encounter obstacles (analogous to antigen), they communicate with software agents (B-cells) residing in the local computer server (LN).  The software agents transmit actions to robots to help overcome their obstacles, and agents also share information globally by migrating to other computer servers (analogous to B-cells migrating between LN). Hence computer servers share efficient solutions locally with robots in their DR and globally with other computer servers. The system is diagrammed in Fig. 2 (modified from \cite{13}).

\subsubsection{Modifying the Original System Using the Modular RADAR Architecture}
By analogy with the NIS, a modular architecture can be used to minimize the time taken by a robot to transmit information about an obstacle (local detection), the time taken by a computer server to transmit back an initial rule-set of actions (local response) and the time taken by a computer server to communicate good rule-sets to other agents (global response). There are two potential communication bottlenecks: communication between robots and computer servers, and communication between computer servers. 

A bottleneck in (local) communication between robot and server demands many small DRs. A bottleneck in (global) server communication requires a few large servers. If both local and global communication are constrained, the architecture which balances these opposing requirements is sub-modular, i.e. the number of servers increases sublinearly with system size and the capacity of each server (bandwidth, memory and number of robots serviced by each server) increases sublinearly with system size (shown in Fig. 2).

\begin{figure}[ht!]
 \includegraphics[width=1\textwidth]{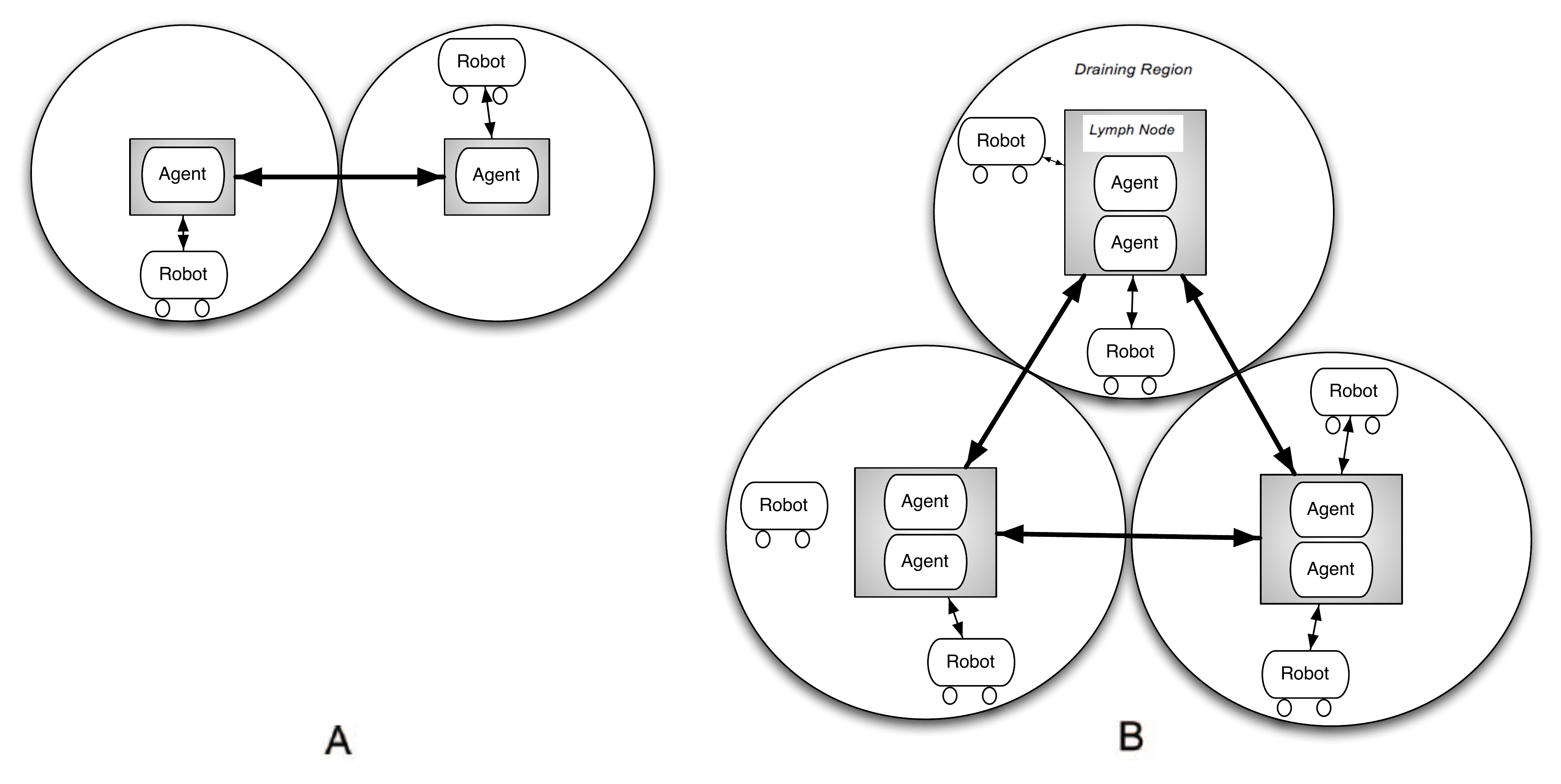}
\caption{(A) Left Panel: a scaled down version of the multi-robot AIS system. The shaded regions are artificial LNs (computer servers) and the unshaded regions are the artificial DR. Light arrows denote communication between robots and servers (local communication) and bold arrows denote communication between servers (global communication). (B) Right Panel: a scaled up multi-robot AIS system with sub-modular architecture. Note that the number of artificial LNs and their size (the number of robots they service and the number of software agents they have in memory) both increase with the size of the system.} 
 \label{fig:Fig2}
\end{figure}

The local communication time within an artificial DR is a function of the number of robots ($d$) serviced by the artificial LN 
\begin{equation}
t_{local} = f(d)
\end{equation}
 
The function $f$ will depend on constraints on communication between robots and servers, influenced, for example, by how robot requests are queued on the server and the distance over which low power robots can send and receive messages. The global communication time between artificial LNs is also a function of the number of LNs in the system ($n/d$) where $n$ is the total number of robots in the entire system

\begin{equation}
t_{global} = g(n/d)
\end{equation}

The function $g$ depends on communication constraints between servers. For low latency and high bandwidth connections among servers, $t_{global}$ may not scale appreciably. However, low power servers distributed in remote environments, may preclude broadcast communication such that $t_{global}$ increases with $n/d$.

An increase in the size of an artificial LN (and hence the number of robots serviced, $d$) would reduce $t_{global}$ at the cost of $t_{local}$. The size and number of artificial LNs to balance local and global communication depends on the precise functions $f$ and $g$ mediating local and global communication. We turn to another application in peer-to-peer systems where local and global communication are specified.  

In summary, the incorporation of a modular RADAR strategy will lead to faster search and response times ($T = t_{local} + t_{global}$) in such LN inspired AIS applications. It would also improve robustness since there are more components (robots in this case) in a single DR as the system grows larger, and hence failure can be compensated by the presence of other redundant components. 

\subsection{Peer-to-Peer Systems}

Peer-to-peer (P2P) systems are emerging as a significant vehicle for providing distributed services like search, content integration and administration. In such systems, computer nodes store data or a particular service and no single node has complete global information about the whole system. Hence these decentralized systems present fundamental challenges during location of resources (data, services, etc) distributed over a multitude of locations.

\subsubsection{Original System}
Search for resources (data, services, etc) distributed over multiple nodes in a P2P network, is similar to the search for pathogens by DCs within a LN and the search for cognate B-cells between LN. Here we focus on a specific P2P overlay network called Semantic Small World (SSW) that supports efficient context-semantic based search \cite{16}. SSW represents objects by a collection of attribute values derived from object content or metadata. The SSW P2P overlay network aggregates data objects with similar semantics close to each other in clusters in order to facilitate efficient search.

Real-world applications require a large number of attributes to identify data objects, so search is through a high-dimensional search space. SSW follows Kleinberg \cite{17} to understand how to conduct efficient decentralized search when each node only has information about its neighbours, and no nodes have global knowledge about the location of resources. For such search to efficiently scale up to large numbers of nodes, each node is required to maintain some long-distance connections drawn from a particular probability distribution. Each node maintains $l$ long-distance links each of which are drawn with probability proportional to $1/d$ ($d$ is the distance between two nodes) in addition to $s$ short-distance links. With such a \textit{small-world} distribution of links, the network structure itself provides latent structural cues, such that each node with only local information can guide a message to a distant target \cite{17}. 

Search across clusters (global search) proceeds via the short and long distance links by comparing coordinates of the destination and subspaces of the traversed nodes. Search within a cluster (local search) is by neighbour flooding whereby each node sends out a search request to all its nearest neighbours. The system is diagrammed in Fig 3. The average time for search across clusters (global search) is given by
$t_{global} = O(\frac{log^{2}(n/c)}{l})$ where $n$ is the number of nodes, $c$ is the size of a cluster (number of nodes in a cluster) and $l$ is the number of long distance connections per node \cite{16}. 

\subsubsection{Modifying the Original System Using the Modular RADAR Architecture}

SSW keeps the size of a cluster ($c$) and the number of long-distance links per node ($c$) constant, even as the number of nodes, $n$, increases. Our contribution is to modify the system by varying the cluster size and the number of long-distance links per node, analogous to the way that the NIS varies LN size and communication between LN, respectively. We show that there exists a tradeoff between local search time (within a cluster) and global search time (across clusters), and the modular RADAR architecture efficiently balances them. 

\begin{figure}[ht!]
 \includegraphics[width=1\textwidth]{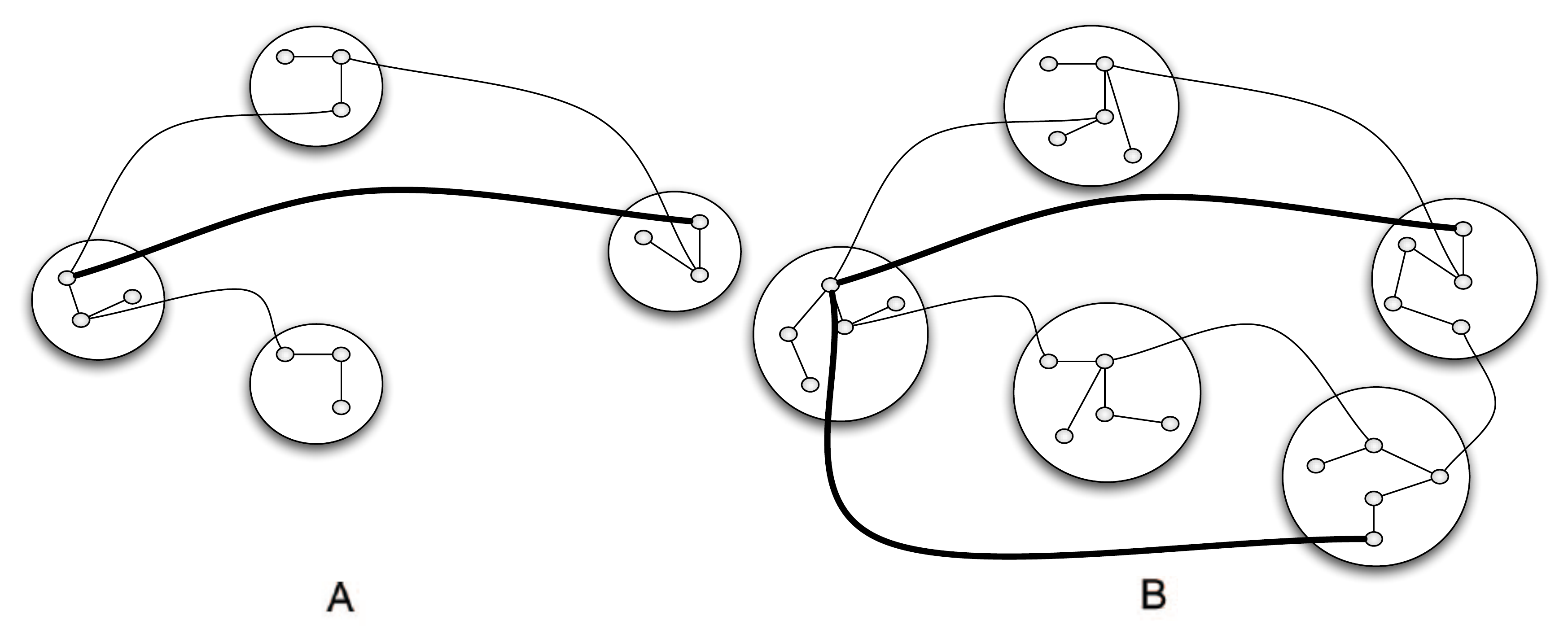}
\caption{(A) Left Panel: a P2P content search system showing clusters (big circles) and nodes within them (small ovals). Short-distance links are shown by light arrows and a long-distance connection is shown in bold arrows (shown only for a single node). (B) Right Panel: a scaled up version of the P2P content search system with a sub-modular architecture. Both the number of clusters and their size (number of nodes within them) now increase with system size (total number of nodes). Note that there are also more long-distance connections per node (shown only for a single node).} 
 \label{fig:Fig3}
\end{figure}

First we vary the number of long-distance connections per node as $l = O(log(n/c))$. This sets the number of long-distance links per node to the logarithm of the number of clusters ($n/c$). This is an example of \textit{densification} \cite{18} in which the number of connections increases slowly with system size in order to minimize the time to send a message to a distant node and also increases robustness. Densification is found empirically in technological networks \cite{18}. The resulting average time for search across clusters is now given by $t_{global} = O(log(n/c))$. The worst case time taken to propagate a message within a cluster (local search) using neighbour flooding is given by $t_{local} = O(c^{1/2})$. This reflects the longest message propagation distance (the diagonal) in a lattice of $c$ elements.
Thus, the total time to search within and across clusters is given by
\begin{equation}
T = t_{local} + t_{global} =  \alpha_{1}c^{1/2} +\alpha_{2}log(n/c)
\end{equation}
where $\alpha_{1}$ and $\alpha_{2}$ are constants. Simplifying we have $T = \alpha_{1}c^{1/2} - \alpha_{2}log c + \alpha_{2}log n$. Since we desire a relation for $c$ (cluster size) in terms of $n$ (number of nodes), intuitively we see that the expression containing $c$ ($\alpha_{1}c^{1/2} - \alpha_{2}log c$) must scale at most as the term containing $n$ ($\alpha_{2}log n$). Otherwise, the expression containing $c$ asymptotically dominates the expression containing $n$, and $T$ would scale faster than $log n$. To achieve better than logarithmic scaling with $n$, we set $\alpha_{1}c^{1/2}$ and $\alpha_{2}log n$ equal to each other giving $c = O(log^{2}n)$. Making these notions concrete we minimize $T$ by differentiating with respect to $c$ and have
\begin{equation}
\frac{dT}{dc} = \frac{\alpha_{1}}{2c^{1/2}} - \frac{\alpha_{2}}{n} \frac{dn}{dc} - \frac{\alpha_{2}}{c} = 0
\end{equation}
Simplifying and integrating we have $\alpha_{2}log c - \alpha_{1}c^{1/2} = \alpha_{2}log n + F$ where the constant $F$ subsumes all the constants of integration. Asymptotically the optimal cluster size
\begin{equation}
c = O(log^{2}n)
\end{equation}
The resulting total time for search in the SSW overlay network is now
\begin{equation}
T = O(logn - loglogn)
\end{equation}

Intuitively a larger cluster size reduces the number of clusters leading to a reduction in global search time across cluster. However a larger cluster size also increases local search time since there are more nodes. Our architecture balances these tradeoffs and improves the scaling of search time ($T$) over prior P2P systems \cite{19}. In addition to achieving rapid search, the modified P2P system is also robust and adaptive. The system is robust to node failures since there is a cluster of size $O(log^{2}n)$ leading to increased robustness by redundancy of similar data. Additionally, if search for data residing on a particular node does not succeed due to node failure, data similar to the one being searched for can be supplied by neighbouring semantically clustered nodes. Each node has connections with multiple clusters which leads to more redundant search pathways.

The system adapts by having progressively more long-distant connections ($l = O(logn)$) in order to improve its search characteristics. Such densification properties are also observed in other technological networks (like the World Wide Web, citation networks, and autonomous systems) \cite{18} and arise as an emergent property of these systems. The cluster size also adaptively changes to efficiently balance local and global search performance.

In summary, we see that similar to an NIS, P2P systems exhibit tradeoffs between local communication (search within a cluster of nodes) and global communication (search across clusters). A sub-modular architecture in which the size of a cluster increases with system size is shown to efficiently balance these opposing goals. We modified an existing P2P system by incorporating a modular RADAR search strategy inspired by the natural immune system and demonstrated that it can conduct efficient search while being robust and adaptable.

\section{Future Work}

Future work will focus on augmenting two intrusion detection applications - LISYS \cite{20} and process Homeostasis (pH) \cite{21}, with a modular RADAR strategy. In this context, a subnet of computers would be an artificial DR and a security node in charge of the subnet would be analogous to a LN. A modular RADAR architecture can balance local communication (intrusion detection) and global communication (alert or patch propagation). This would further advance work on the architecture of intrusion detection systems \cite{22}.

Such an approach may be particularly useful for security of low power distributed systems. For example,  mobile phone viruses propagate by small hops to neighbouring within-range devices, and hence physical proximity of devices is much more important than in traditional computational domains \cite{1}. Mobile phone viruses also consume low power and bandwidth in order to evade conventional detection mechanisms. Since mobile phones are constrained to communicate to nearby neighbors (e.g. through Bluetooth connections and local cell towers), the NIS analogy can be extended to view mobile phone transmission towers as LNs and the area of mobile phone users serviced by it as the DR. 

Finally, mobile ad-hoc networks (MANETs) and disruption tolerant networks (DTNs) \cite{1} are of increasing interest. Such systems could also have a master-slave relationship in which individual components report aggregated data to a processing center (local communication) and processing centers distribute data globally among all components (global communication) \cite{23}. A modular RADAR strategy can enhance message propagation times and increase robustness in such systems.

In all of these systems, the balance between LN size and number depends on the constraints on local and global communication. In the NIS, we considered one scenario in which the pathogen is initially localized in the tissue of a single DR, and the draining LN must recruit  additional B-cells from other LN through HEV that deliver blood. However this is one of several scenarios that the IS architecture has evolved. For example, some pathogens propagate immediately through blood and are exposed to multiple LN at once. In such a scenario there is no need for recruitment to a single LN because the pathogen is distributed to multiple LNs. In other scenarios, an infection may stay local, and again there is no need for recruitment. Understanding how the semi-modular architecture of the immune system balances these different scenarios is another area for future work. 

\section{Conclusions}

The physical \textit{architecture} of the immune system has co-evolved with the \textit{software}, i.e. dynamic interactions between immune system cells and signals to achieve scalable Robust Adaptive Decentralized search and Automated Response (RADAR). In this paper we have focused on how and why search is distributed across LN, whose size and number increase sublinearly with animal size. We have shown that such a design balances local antigen detection with global antibody response to minimize total immune response time. As a result, immune response times are nearly \textit{scale invariant}, in sharp contrast to the vast majority of biological rates that slow systematically as body size increases \cite{24}. We have suggested that modular design principles from the immune system can complement the AIS software approaches to distributed computation, particularly to achieve scalable mobile device control, peer-to-peer networks, and potentially computer security applications.

\section{Acknowledgements}
We would like to acknowledge fruitful discussions with Dr. Alan Perelson, Dr. Stephanie Forrest and Dr. Jedidiah Crandall. This work supported by a grant from the National Institute of Health (NIH RR018754). S.B. would like to acknowledge travel grants from RPT, SCAP and PIBBS at the University of New Mexico.

\bibliographystyle{splncs03}

\end{document}